\documentclass[aps,prb,showpacs,superscriptaddress,twocolumn,amsmath,amssymb]{revtex4-1}

\usepackage{graphicx}
\usepackage{dcolumn}
\usepackage{bm}
\usepackage{ifthen}
\usepackage{booktabs}
\usepackage{SIunits}
\usepackage{ulem}
\usepackage{xcolor}
\newcommand*{\bra}[1]{\langle #1 |}
\newcommand*{\ket}[1]{| #1 \rangle}

\begin{document}


\title{Theory of strain tunning exction coupling in self-assembled
  InAs/GaAs quantum dots}

\author{Jianping Wang}
\affiliation{Key Laboratory of Quantum Information, University of Science and
Technology of China, Hefei, 230026, China}
\affiliation{Synergetic Innovation Center of Quantum Information and Quantum
  Physics, University of Science and Technology of China, Hefei, 230026, China}
\author{Guang-Can Guo}
\affiliation{Key Laboratory of Quantum Information, University of Science and
Technology of China, Hefei, 230026, China}
\affiliation{Synergetic Innovation Center of Quantum Information and Quantum
  Physics, University of Science and Technology of China, Hefei, 230026, China}
\author{Lixin He}\email{helx@ustc.edu.cn.}
\affiliation{Key Laboratory of Quantum Information, University of Science and
Technology of China, Hefei, 230026, China}
\affiliation{Synergetic Innovation Center of Quantum Information and Quantum
  Physics, University of Science and Technology of China, Hefei, 230026,
  China}

\date{\today}

\begin{abstract}

We derive analytically the change of exciton fine structure splitting (FSS)
 under the external stresses in the
self-assembled InAs/GaAs quantum dots using the
Bir-Pikus model. We find that the FSS change is mainly due to the
strain induced valence bands mixing and valence-conduction band coupling. The
exciton polarization angle under strain are determined by the argument of the
electron-hole off-diagonal exchange integrals.  The theory agrees well with the
empirical pseudopotential calculations.

\end{abstract}

\maketitle

\section{Introduction}

Self-assembled quantum dots (QDs), also known as artificial atoms, are made of
millions of atoms. \cite{bimberg_book} Unlike real atoms, which have constant
physical properties, the QDs are different from each other and show far more
complicate behavior.  The physical properties of QDs, are determined by 
the combined effects of the strain distributions, alloy, interface etc., 
which are coined during their growing process. It is a great challenge, as well as an
opportunity, to tune the QDs to desired properties (e.g., the exciton energy,
polarization, and fine structure splitting etc.) by external fields, which is
not only interesting for fundamental physics, but is also extremely important
for device applications.  However, despite the importance, our understanding to
the interplay between the QDs and external fields is still very limited.

One of the most prominent applications of QDs is as the entangled photon 
emitters, based on the biexciton cascade process, \cite{benson00,stevenson06} 
which has attracted enormous interest in the last decade. 
However, though it is simple in principle, it is not easy to
implement experimentally.
This is because of the existence of in-plane anisotropy in the QDs, the
two biexciton decay pathways may have a small energy difference known as the fine
structure splitting (FSS). When the FSS exceeds the radiative linewidth ($\sim$
1.0 $\mu$eV), the polarization entanglement will be destroyed.\cite{stevenson06,gong08}  
Great effort has been made to reduce the FSS using various post-growth tuning
techniques.\cite{bennett10,gerardot07,vogel07,trotta12,ding10,jons11,seidl06,dou08,gong11}
Especially, it has recently been demonstrated that the FSS can be universally
suppressed through the combination of electric field and
stresses,\cite{wang12,trotta12} 
regardless of the dots' details.

In previous works, we have developed an effective model~\cite{gong11,wang12}  
based on symmetry analysis to explain how the FSS change 
under external stress. It turns out that the results obtained 
from the simple effective model is in excellent agreement with 
those obtained from a more sophisticated empirical
pseudopotential method (EPM) \cite{williamson00} 
and configuration interaction (CI) calculations \cite{Franceschetti1999} 
and as well as the experimental results. \cite{plumhof11,kuklewicz12,trotta12} 
However, there is a hug gap between 
the effective model and the EPM calculations 
in understanding how exactly the strain modify the exciton coupling
in the QDs at microscopic level. 

The purpose of this paper is to bridge the gap between the effective model and
the pseudopotential calculations. We derive analytically the exciton FSS 
under the external stresses in self-assembled InAs/GaAs QDs
using the Bir-Pikus model.\cite{pikus} We show that the strain
induced valence bands mixing and valence-conduction bands (VB-CB) coupling 
play the most important roles in tuning the FSS. 
Detailed comparison between the Bir-Pikus model and the
EPM calculations shows that the simple Bir-Pikus model provides
semi-quantitatively description of the FSS under strain. We further clarify
the polarization angle change under the external stresses.

The rest of paper is organized as follows. In Sec. \ref{sec:single-particle},
we discuss how the single-particle states of a QD vary under the external
stresses using the Bir-Pikus model.
We discuss how the electron-hole exchange integrals and 
FSS change under the external stresses in Sec. \ref{sec:FSS},
and the exciton polarizations in Sec. \ref{sec:polarization}.
We summarize in Sec. \ref{sec:summary}.

\section{Single-particle states in a QD under external strain}
\label{sec:single-particle}

We first look at how the single particle states in a QD vary
under external strain field. Usually the applied uniaxial stress is 
less than $\pm$ 100 MPa. 
Under such small stress, the shape of QDs changes very little. We therefore 
neglect the change of envelope functions of the single particle states, and
focus on the underlying atomistic wave functions. We further
assume that dots have uniformly distributed strain due to the lattice mismatch
between the InAs dot and GaAs matrix, and neglect the interface effects for
the moment.

The influence of strain on valence states in zinc-blende 
structures can be described by the Bir-Pikus model.\cite{pikus}
We expand the Bir-Pikus Hamiltonian with in the six $\ket{j,j_z}$ states, i.e.,
heavy hole (HH) $\ket{3/2,\pm 3/2}$ , light hole (LH)  $\ket{3/2,\pm 1/2}$ 
and spin orbital (SO) $\ket{1/2,\pm 1/2}$ states, resulting in
the following $6\times6$ matrix,
\begin{equation}\label{eq:HBP}
\left(\begin{array}{cccccc}
P+Q               & 0          & -\sqrt{2}S         & R                 & S                & \sqrt{2}R \\
0                 & P+Q        & R^{\ast}           & \sqrt{2}S^{\ast}  & \sqrt{2}R^{\ast} & -S^{\ast} \\
-\sqrt{2}S^{\ast} & R          & P-Q                & 0                 & \sqrt{2}Q        & -\sqrt{3}S\\
R^{\ast}          & \sqrt{2}S  & 0                  & P-Q               & \sqrt{3}S^{\ast} & \sqrt{2}Q \\
S^{\ast}          & \sqrt{2}R  & \sqrt{2}Q          & \sqrt{3}S         & P                & 0         \\
\sqrt{2}R^{\ast}  & -S         & -\sqrt{3}S^{\ast}  & \sqrt{2}Q         & 0                & P
\end{array}\right)\, ,
\end{equation}
where,
\begin{eqnarray}
P &=& a_{v}(e_{xx}+e_{yy}+e_{zz}) \, ,\\ \nonumber 
Q &=& \frac{b_v}{2}(e_{xx}+e_{yy}-2e_{zz}) \, , \\ \nonumber 
R &=& \frac{\sqrt{3}}{2}b_v(e_{xx}-e_{yy})-id_v e_{xy}\, ,\label{eq:BP-R} \\ \nonumber
S &=& \frac{d_v}{\sqrt{2}}(e_{zx}-ie_{yz})\, .
\label{eq:PQRS}
\end{eqnarray}
$a_{v}$, $b_{v}$, and $d_{v}$ are the isotropic, biaxial, and shear
deformation potentials respectively and $e_{ij}$ are the strain components 
in the QDs. $P$ describes the effects of isotropic hydrostatic strain and $Q$ is
associated with the biaxial strain. The effects of in-plane
and off-plane strain anisotropy are accounted by $R$ and $S$.
In self-assembled InAs/GaAs QDs grown on the (001) GaAs substrate,
the dot material is compressed in the growth plane and distended in the growth direction.
We also consider the effects of strain anisotropy in the growth plane ($e_{xy}$,
$e_{xx}-e_{yy}$) and off-plane shear strains ($e_{zx}$ and $e_{yz}$).

For most III-V semiconductors, the SO bands are several hundreds meV
below the HH and LH bands. The SO band were ignored in many previous
works.\cite{Leger2007,Testelin2009,Tonin2012}
However, for self-assembled InGaAs/GaAs QDs and other nano-structures with large
lattice mismatch, the biaxial strain is very large, which push the LH bands
down towards the SO bands, therefore the coupling to the SO band is also
important, as will be demonstrated in this work.
Therefore, the full Hamiltonian should also includes the 
SO coupling term, 
\begin{equation}
H_{\mathrm{SO}}=\frac{\Delta}{3}\left(\begin{array}{cccccc}
1 & 0 & 0 & 0 & 0 & 0 \\
0 & 1 & 0 & 0 & 0 & 0 \\
0 & 0 & 1 & 0 & 0 & 0 \\
0 & 0 & 0 & 1 & 0 & 0 \\
0 & 0 & 0 & 0 & -2& 0 \\
0 & 0 & 0 & 0 & 0 & -2
\end{array}\right).
\end{equation}
Here, $\Delta \sim$ 390 meV is the SO parameter in InAs. 
The total Hamiltonian is given by $H=H_{\mathrm{BP}}+H_{\mathrm{SO}}$.

Because of the large lattice mismatch (7\%) between InAs and GaAs in the InAs/GaAs
QDs, the biaxial strain ($|e_{xx}+e_{yy}-2e_{zz}| \sim$ 24\%) is much larger than
the shear strains ($|e_{xx}-e_{yy}|\sim|e_{yz}|\sim|e_{zx}|\sim$ 1 \%, $|e_{xy}|\sim0.5$\%). 
As a consequence, $Q$ is comparable with the
SO parameter $\Delta$, much larger than $|R|$ and $|S|$.
Therefore, we treat $R$ and $S$ as perturbations in the Hamiltonian.
We calculate the first two (degenerate) hole states states up to 
second order of $R$ and $S$,
\begin{eqnarray}
   \mathcal{N}\ket{\psi^{v}_{-}} =
   \ket{\frac{3}{2},+\frac{3}{2}} 
&+\chi_{\alpha}\ket{\frac{3}{2},+\frac{1}{2}}
+\chi_{\beta}\ket{\frac{1}{2},+\frac{1}{2}} \nonumber\\
&+\varepsilon_{\alpha}\ket{\frac{3}{2},-\frac{1}{2}}
+\varepsilon_{\beta}\ket{\frac{1}{2},-\frac{1}{2}},
 \nonumber \\
\mathcal{N}\ket{\psi^{v}_{+}}= \ket{\frac{3}{2},-\frac{3}{2}}
&+\varepsilon_{\alpha}^{\ast}\ket{\frac{3}{2},+\frac{1}{2}}
+\varepsilon_{\beta}^{\ast}\ket{\frac{1}{2},+\frac{1}{2}}
\nonumber\\&-\chi_{\alpha}^{\ast}\ket{\frac{3}{2},-\frac{1}{2}}
-\chi_{\beta}^{\ast}\ket{\frac{1}{2},-\frac{1}{2}},
\end{eqnarray}
where
\begin{eqnarray}
  \varepsilon_{\alpha}&=&\frac{(3Q+\Delta)R^{\ast}+\sqrt{3}(S^{\ast})^2}{2Q\Delta}
  \, ,\nonumber \\
  \varepsilon_{\beta}&=&\frac{3\sqrt{2}Q
    R^{\ast}+\sqrt{6}(S^{\ast})^2}{2Q\Delta} \, ,\nonumber\\
  \chi_{\alpha}&=&\frac{-\sqrt{2}\Delta S^{\ast}-\sqrt{6}SR^{\ast}}{2Q\Delta}
  \, ,\nonumber\\
  \chi_{\beta}&=&\frac{\sqrt{3}SR^{\ast}}{2Q\Delta}\, ,
\label{eq:mixing}
\end{eqnarray}
and $\mathcal{N}^2=1+|\varepsilon_{\alpha}|^2+|\varepsilon_{\beta}|^2+|\chi_{\alpha}|^2+|\chi_{\beta}|^2
\sim$ 1 is the normalization factor.
The single-particle energy of the two states is,
\begin{eqnarray}
E_{v} &=& P+Q+{\Delta \over 3}
+{2\Delta|S|^2+(9Q+\Delta)|R|^2 \over 2Q\Delta} \\ \nonumber
&& +\frac{3\sqrt{3}(S^2R^{\ast}+(S^{\ast})^2R)}{2Q\Delta}\, .
\end{eqnarray}
These two states are still dominated by the HH ($j=3/2$) states but mixing up
with some LH and SO components.
There are are two mixing mechanisms:
(i) The mixing between HH and $j_{z}$ states of the opposite signs 
is mainly due to the in-plane anisotropic strain effects or shape asymmetry ($R$).
The mixing amplitude is determined by $\varepsilon_{\alpha}$ and $\varepsilon_{\beta}$.
(ii) The mixing between HH and 
$j_{z}$ states with the same sign is mainly 
due to the off-plane shear strain components ($S$).
The mixing amplitude is determined by
$\chi_{\alpha}$ and $\chi_{\beta}$.
Both mixing  mechanisms have important influence on the optical properties of
the QDs, which will be discussed later in the paper.

For the simplicity of the discussion, we ignore the 
VB-CB coupling for a moment. But we will see later that the VB-CB
coupling is also important for the FSS change under stain, which is
addressed in the appendix.
The Bloch parts of the conduction states are dominated by the lowest electron
bands, $\ket{\psi^{c}_{+}}=\ket{e\uparrow}$ and
$\ket{\psi^{v}_{-}}=\ket{e\downarrow}$.
The energy of conduction states merely depends on hydrostatic strain:
\begin{equation}
\delta E_c (\tensor{e}) = a_c (e_{xx}+e_{yy}+e_{zz}).
\end{equation}

Under external stresses, the strain distribution in QDs
changes accordingly, which changes the single-particle energy levels, 
as well as the coupling between the HH, LH and SO bands. 
We take the uniaxial stress along the [110] direction for example.
The relation between the change of strain and stress $p$ along the
  [110] direction is given by,
\begin{equation}\begin{split}
\Delta e_{xx} & = \Delta e_{yy} = -\frac{1}{2} (S_{11}+S_{12}) p \, ,\\
\Delta e_{zz} & = -2S_{12} p\, , \\
\Delta e_{xy} & = -\frac{1}{4} S_{44} p \, ,\\
\Delta e_{zx} & = \Delta e_{yz} = 0\, .
\end{split}
\end{equation}
Here, we take the compressive stress as positive one,
and the parameters $P$, $Q$, $R$, $S$ in the
Bir-Pikus Hamiltonian under stress along the [110] direction 
can be written as,
\begin{eqnarray}
P(p) & =& P(0) - a_{v} (S_{11}+2 S_{12}) p \, , \nonumber \\
Q(p) & =& Q(0) - \frac{1}{2} b_{v} (S_{11}-S_{12}) p \, \nonumber \\
R(p) & =& R(0) + i \frac{1}{4} d_{v} S_{44} p \, \nonumber \\
S(p) & =& S(0) \, .
\label{eq:pqrs}
\end{eqnarray}
Interestingly, $S$ does not change with  
the stress along the [110] direction. 
We have
\begin{equation}
\frac{d E_v}{d p} \approx -a_{v} (S_{11}+2 S_{12})-\frac{1}{2} b_{v}
(S_{11}-S_{12}) \, ,
\end{equation}
and
\begin{equation}
{dE_{c} \over d p}= -a_c (S_{11}+2S_{12}).
\end{equation}
Because the envelope functions of the electron and hole states change little,
if the external stress is not very large, the direct electron-hole 
Coulomb interaction also change little. 
The change of the exciton energy is therefore mainly determined by
the single-particle energies. We can estimate the of energy change to the
stress along [110] direction as,
\begin{equation}
\frac{d E_{X^{0}} }{d p} \approx -(a_c - a_v)(S_{11}+2S_{12})+\frac{1}{2}b_{v}
(S_{11} - S_{12})\, .
\label{eq:exciton-energy}
\end{equation}
Using the deformation potential parameters for bulk InAs material and elastic
compliance constants for bulk GaAs material listed in Table \ref{tab:param},
we obtain $d E_{X^{0}}/d p \approx$ 12.3 $\mu$eV/MPa. This value is in consistent with
recent experimental results. \cite{kuklewicz12}
Although the exciton energy can be tuned by the stress
along [110] direction, the tuning slope 
is rather small, because of the cancellation effect between the conduction 
band and valence bands in Eq. \ref{eq:exciton-energy}. Furthermore, in QDs
the confinement potentials, alloy effects, etc. may
also plays important roles to the exciton emission energies, 
therefore, ${d E_{X^{0}} / d p}$ may vary from
dots to dots. \cite{kuklewicz12}
 
\begin{table}[htbp]
\caption{\label{tab:param} The compliance constants 
for GaAs and and deformation potentials for InAs. 
The deformation potentials for strained InAs is calculated by EPM using
isotropic and biaxial strains for typical QDs.}
\centering
\begin{tabular}{lcccc}
\hline\hline
Parameters & Unit & GaAs & InAs(bulk)\cite{Vurgaftman2001}  & InAs (strained)\\
\hline
$a_{c}$    & eV   & -- & -5.08 & --\\
$a_{v}$    & eV   & -- & -1.0  & -0.23\\
$b_{v}$    & eV   & -- & -1.8  & -2.22\\
$d_{v}$    & eV   & -- & -3.6  & -6.49\\
$\Delta$& eV & -- & 0.39  & 0.33\\
$S_{11}$   & 10$^{-2}$ GPa$^{-1}$ &  1.17 & &\\
$S_{12}$   & 10$^{-2}$ GPa$^{-1}$ & -0.37 & &\\
$S_{44}$   & 10$^{-2}$ GPa$^{-1}$ &  1.68 & &\\
\hline\hline
\end{tabular}
\end{table}

\begin{figure}
\includegraphics[width=3.4in]{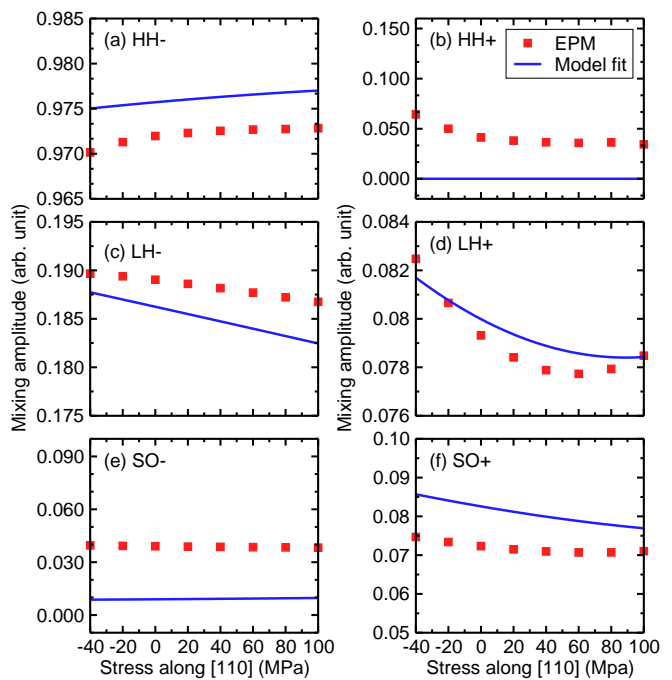}
\caption{\label{fig:mixing} (Color online)
Valence bands mixing of the first
hole state in a pure 
lens-shaped InAs/GaAs dot with base $D$=25 nm and height
$h$=3.5 nm under uniaxial stress along the [110] direction, showing
(a) HH$_+$, (b) HH$_-$, (c) LH$_+$, (d)
LH$_-$, (e) SO$_+$ and (f) SO$_-$ components. The red squires are calculated
from EPM, whereas the blue lines are obtained from Bir-Pikus model.}
\end{figure}

\section{Electron-hole exchange interaction and FSS}
\label{sec:FSS}

In this section we discuss how the external strain modifies the exciton exchange energies
and the FSS.
The matrix elements of the exciton Hamiltonian between different spin 
configurations is written as,~\cite{Franceschetti1999}
\begin{equation}\begin{split}
\mathcal{H}_{v'c',vc}&=\bra{\Phi_{v'c'}}\mathcal{H}\ket{\Phi_{vc}}
\\&=(E_c-E_v)\delta_{c,c'}\delta_{v,v'}
-J_{v'c',vc}+K_{v'c',vc} \, ,
\end{split}\end{equation}
where the $J$s and $K$s are the Coulomb and exchange integrals respectively.
We consider only the first two hole states ($\psi^{v}_{+}$ and $\psi^{v}_{-}$) and
electron states ($\psi^{c}_{+}$ and $\psi^{c}_{-}$).
According to the symmetry, the only configurations with anti-parallel spins
$(\psi^{v}_{-} \psi^{c}_{+}, \psi^{v}_{+} \psi^{c}_{-})$ make contributions to bright excitons (BE).
In this basis, the many-particle Hamiltonian for bright excitons is
\begin{equation}
H_{\textrm{BE}}=(E_c-E_v)-J_{eh}+\begin{bmatrix}
	K_{\rm d} & K_{\rm od} \\
	K^{\ast}_{\rm od} & K_{\rm d} \end{bmatrix}\, , \\
\label{eq:exciton}
\end{equation}
where $J_{eh}$ is the electron-hole Coulomb interaction, $K_{\rm d}$ is the
diagonal exchange energy, which determines the dark-bright exciton energy
splitting, whereas the off-diagonal exchange energy, 
\begin{equation}\begin{split}
K_{\rm od} &= \bra{\psi^{v}_{-}\psi^{c}_{+}}\mathcal{K}_{\mathrm{ex}}\ket{\psi^{v}_{+}\psi^{c}_{-}},
\\ & = \iint\frac{[\psi_{+}^{v}(x_1)\psi_{+}^{c}(x_2)]^{\ast}\psi^{c}_{-}(x_1)\psi^{v}_{-}(x_2)}
    {\bar{\mathcal{\epsilon}}(\mathbf{r}_1,\mathbf{r}_2)|\mathbf{r}_1-\mathbf{r}_2|}
    dx_1dx_2 \, ,
\end{split}\end{equation}
is responsible for the bright exciton energy splitting. 
After diagonalization of Eq.~\ref{eq:exciton}, 
the eigenstates of the two bright exciton 
can be written as,
\begin{align}
\ket{B_{1}}&= \frac{1}{\sqrt{2}}(\ket{\psi^{v}_{-}\psi^{c}_{+}} +
  \mathrm{e}^{i2\theta}\ket{\psi^{v}_{+}\psi^{c}_{-}}) \, ,\\
\ket{B_{2}}&= \frac{1}{\sqrt{2}}(\ket{\psi^{v}_{-}\psi^{c}_{+}} -
  \mathrm{e}^{i2\theta}\ket{\psi^{v}_{+}\psi^{c}_{-}})\, ,
\end{align}
with 2$\theta=-\mathrm{arg}(K_{\rm od})$.
The energy splitting between the two bright excitons,
which is known as FSS, is given by $\Delta_{\mathrm{FSS}}=2|K_{\rm od}|$.
Let $a=(X+iY)/\sqrt{2}$ and $b=(X-iY)/\sqrt{2}$, and since
only the hole and electron components of opposite spins in the same configuration
have none-zero contribution to the exchange integral, the exchange integral can be
written as (drop the spin index),
\begin{equation}
K_{\rm od}
={\mathcal{N}}^{-1}\bra{(a +\varepsilon_+ b+\chi_- Z) e}\mathcal{K}_{ex}
\ket{(b +\varepsilon_+^{\ast} a +\chi_-^{\ast} Z) e} \, ,
\end{equation}
with
\begin{align}
  \varepsilon_+&=\frac{\varepsilon_{\alpha}+\sqrt{2}\varepsilon_{\beta}}{\sqrt{3}}
  =\frac{R^{\ast}}{2\sqrt{3}}\left(\frac{1}{Q}+\frac{9}{\Delta}\right)
  +\frac{3(S^\ast)^2}{2Q\Delta} \, , \nonumber \\ 
\chi_-&=\frac{-\sqrt{2}\chi_{\alpha}+\chi_{\beta}}{\sqrt{3}}
  =\frac{S^{\ast}}{\sqrt{3}Q}+\frac{3SR^\ast}{2Q\Delta} \, .
\label{eq:epsilon}
\end{align}
To simplify the notation, we introduce the following parameters,
\begin{align}
\bra{a e} \mathcal{K}_{ex} \ket{a e}
    \equiv\bra{b e} \mathcal{K}_{ex} \ket{b e} &=  K , \nonumber\\
\bra{a e} \mathcal{K}_{ex} \ket{b e} &= \kappa+i\delta , \nonumber \\
\bra{a e} \mathcal{K}_{ex} \ket{Z e}
    \equiv\bra{Z e} \mathcal{K}_{ex} \ket{b e} &=  \mu+i\nu ,\nonumber\\
\bra{Z e} \mathcal{K}_{ex} \ket{Z e} &= K_{z}\label{eq:integral}\, .
\end{align}
Each parameter appearing in Eq.~(\ref{eq:integral}) can be expressed
as exchange integrals over different orbital functions ($X$, $Y$, $Z$, $S$).
For simplicity, we choose all orbital functions to be real. Therefore
the parameters introduced here are all real. Exchange integrals over heavy holes
2$K$ is approximately the dark-bright splitting, and
\[ \kappa=\frac{1}{2}(\bra{Xe} \mathcal{K}_{ex} \ket{Xe} - \bra{Ye}
\mathcal{K}_{ex} \ket{Ye}), \]
comes from the non-equivalence the orbital wave functions $X$ and $Y$, whereas, 
\[ \delta=\frac{1}{2}(\bra{Xe} \mathcal{K}_{ex} \ket{Ye}+\bra{Ye}
\mathcal{K}_{ex} \ket{Xe})\, , \]
is due to the non-orthogonality between the orbital functions $X$, and
$Y$. $\mu$ and $\nu$ is due to the non-orthogonality between the orbital
functions $X$, $Y$ to $Z$,
\begin{align}
 \mu&=  \frac{1}{\sqrt{2}}\bra{Xe} \mathcal{K}_{ex} \ket{Ze} \, , \nonumber \\
\nu &=  \frac{1}{\sqrt{2}}\bra{Ye} \mathcal{K}_{ex} \ket{Ze}\, .
\end{align}
With the above parameters, the whole exchange integral can be written as,
\begin{equation}
\begin{split}
K_{\mathrm{od}}=&
\frac{1}{\mathcal{N}} [ (\kappa+i\delta) +2\varepsilon_+ K +2\chi_- (\mu+i\nu)
\\ &+2\varepsilon_+\chi_- (\mu-i\nu) +\varepsilon_+^2 (\kappa-i\delta)
+\chi_-^2 K_{z} ]\, .
\end{split}
\label{eq:exchange}
\end{equation}
One can see clearly from Eq.~(\ref{eq:exchange}), the origin of the FSS 
from the microscopic structure in the self-assembled
QDs, apart from the dot shape asymmetry, including the 
non-orthogonality and non-equivalence between the atomic orbitals and the band
mixing:  
 
(i) For ideal QDs with $D_{2d}$ and $C_{4v}$ symmetry (e.g., a pure InAs/GaAs quantum disk),
$e_{xx}-e_{yy}$=0, and $e_{xy}$, $e_{yz}$, $e_{zx}$=0, 
there is no coupling between HH with LH and SO bands, and $\kappa$,
$\delta$ also vanish. There would be no FSS.

(ii) For QDs with $C_{2v}$ symmetry (e.g., a pure lens-shaped InAs/GaAs QD), 
the orbital functions $X$ and $Y$ are of mirror symmetry about the [110]
plane, therefore, $\kappa=0$ and $\mu=\nu$.
The strain distribution also obey such mirror symmetry, i.e., 
$e_{xx}=e_{yy}$, $e_{xy}\neq 0$, $e_{zx}=e_{yz}\neq
0$, as a result,  $\varepsilon_+=i|\varepsilon_+|$, $\chi_-=|\chi_-|e^{i\pi/4}$ [See
Eq.~(\ref{eq:PQRS}) and Eq.~(\ref{eq:mixing})]. 
It is easy to verify that $K_{\mathrm{od}}$ 
is pure imaginary.

(iii) For a real dot with $C_1$ symmetry, $K_{\mathrm{od}}$ has both a real
part and an imaginary part.

Because  $|\varepsilon_+|$,$|\chi_-|\ll 1$, 
and $\kappa$, $\delta$, $\mu$, $\nu$ $\ll$ $K$, $K_{z}$,
$K_{\mathrm{od}}$ can be further approximated as $K_{\mathrm{od}}
\approx (\kappa+i\delta) + 2\varepsilon_+ K$.
We assume that the parameters introduced in Eq.~(\ref{eq:integral})
associated with the atomistic orbitals of the underlying dot materials
will not change under small external stress. 
Therefore the change of the exchange integral (away from the critical stress
region \cite{gong11})
can be written as,
\begin{equation}
  {dK_{\mathrm{od}} \over dp} \approx  2\frac{d\varepsilon_+}{dp}K\, .
\end{equation}
Using Eq.~\ref{eq:pqrs} and Eq.~\ref{eq:epsilon}, we have,
\begin{equation}
\frac{d \varepsilon_+}{d p} \approx 
\frac{1}{2\sqrt{3}}\left(\frac{1}{Q}+\frac{9}{\Delta}\right)
\frac{dR^{\ast}}{dp}\, .
\label{eq:dvarepsilon}
\end{equation}
Here, we neglect the change of $Q$ in the denominator. 
For stress applied along the [110] direction,
\begin{equation}
\frac{dR^{\ast}}{dp}=-i\frac{1}{4}d_{v}S_{44}. 
\label{eq:dR}
\end{equation}
Therefore, we have the $K_{\mathrm{od}}$  change under the stress along [110] direction,
\begin{equation}
 {dK_{\mathrm{od}} \over dp} \approx -i
 \left(\frac{9Q+\Delta}{4\sqrt{3}Q\Delta}\right)d_{v}S_{44}K\, .
\label{eq:dK}
\end{equation}
Using parameters given in Table \ref{tab:param}, we get $d\varepsilon_+ / dp =
i\, 2.455 \times 10^{-4} \mathrm{MPa}^{-1}$.  In typical InAs/GaAs QDs, the
exciton dark-bright splitting is approximately $2K\sim$ 200$\mu$eV. Therefore we
estimate that $d\Delta_{\mathrm{FSS}}/dp \sim$ 0.1 $\mu$eV/MPa, which is in the
same order of magnitude with the experimental values\cite{Seidl2006} of
$(0.34\pm0.08) \mu $eV/MPa. As shown in the appendix, the
VB-CB coupling also contribute to the change FSS in a similar magnitude.

Interestingly, as one can see from Eq.~(\ref{eq:dvarepsilon}-\ref{eq:dK}) that
stress along the [110] direction only changes the
imaginary part of the exchange integral $K_{\mathrm{od}}$,
which is just the $\alpha$ parameter defined in Ref. \onlinecite{gong11}.  
It is also easy to
verify that stress along the [100] or [010] direction only changes the real
part of the exchange integral $K_{\mathrm{od}}$, which is $\beta$ defined in 
Ref. \onlinecite{gong11}. (Note that the basis of exciton wave function in
Eq.~\ref{eq:exciton} is different than that used in Ref. \onlinecite{gong11}).
For dots with  $C_{2v}$ symmetry, in which $K_{\mathrm{od}}$ has only an
imaginary part, the stress along the [110] direction alone can tune the FSS to zero,
whereas for dots with $C_1$ symmetry, in which $K_{\mathrm{od}}$ has both a real
part and an imaginary part, the FSS cannot be tuned to zero under single uniaxial stress. 
However, since the stresses along the [110] and [100] directions can manipulate the
imaginary and real parts of $K_{\mathrm{od}}$ (almost) independently, the FSS
can be tuned to nearly zero, as predicted in our previous work.~\cite{wang12}

\begin{figure}
\includegraphics[width=0.4\textwidth]{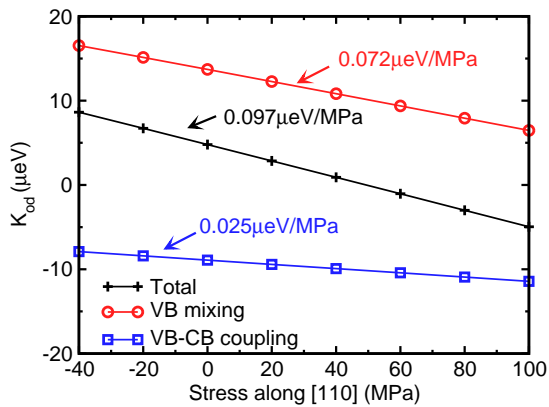}
\caption{\label{fig:Kex}
(Color online) The off-diagonal electron-hole exchange interactions $|K_{\rm od}|$ 
in a lens-shaped InAs/GaAs QD ($D$= 25 nm, $h$= 3.5 nm)
as functions of the uniaxial stress along the [110] direction. The black line
is the total $|K_{\rm od}|$, whereas the red and blue lines are the
contribution from VB mixing and VB-CB coupling respectively.
}
\end{figure}

Of course the above Bir-Pikus model is highly simplified. The real QDs are far
more complicated. To see if this model is valid, we perform EPM calculations on
realistic InAs/GaAs QDs under external stresses.  The dots are assumed to be
grown on the [001] direction and embedded in a $60\times 60\times 60$ GaAs
supercell.  The atom positions in the supercell are optimized by the valence
force field method.\cite{keating66,martin70} We solve the single-particle states
by expanding the wavefunctions with a strained linear combination of Bloch bands
method (SLCBB).\cite{wang99b} The exciton energies are calculated by the CI
method,\cite{Franceschetti1999} in which the exciton wavefunctions are expanded
in Slater determinants constructed from all confined electron and hole
single-particle states.

To compare with the Bir-Pikus model, we project the first hole single-particle
wave function to the $|j, j_z\rangle$ states at $\Gamma$ point. \cite{wei12} The
results are shown in Fig.~\ref{fig:mixing}, compared with those from model
calculations.  The solid squares represent the amplitude of different components
obtained from empirical pseudopotentials calculations by integrating the envelop
functions of each component over the whole supercell. The blue lines are the
results obtained from Bir-Pikus model, using deformation potentials for
strained InAs given in Table \ref{tab:param}.
The wave functions are normalized to 1 in
both cases.  We see that the EPM and Bir-Pikus model results are in a reasonable
good agreement with each other (Note that the scale of the figure is extremely
small).  In the Bir-Pikus model, HH$_+$ and HH$_-$ components do not mix with
each other, whereas in the EPM calculations, there is small mixing of HH$_+$ and
HH$_-$ states, because the SLCBB method use Bloch basis functions from many
$k$-points around the $\Gamma$ point, whereas the Bir-Pikus model~\cite{pikus}
use only the Bloch basis functions at $\Gamma$ point.  Importantly the highly
simplified Bir-Pikus model gives similar slopes of the magnitude of the
components to the external stress as those from atomistic calculations.  The
quantitative differences between the two theories are due to the neglect of the
nonuniform distribution of strain, inter-facial effects, etc. in the Bir-Pikus
model.

We further compare the exchange integrals $K_{\rm od}$ between the two theories.
Figure~\ref{fig:Kex} depicts the EPM calculated exchange integral $K_{\rm od}$ in
a pure lens-shaped InAs/GaAs QD with base $D$=25 nm and height $h$=3.5 nm as a
function of the stress along the [110] direction.  The total $K_{\rm od}$ under
external stress, shown in black line, is $0.097$ $\mu$eV/MPa, and the change
of FSS under stress is 2$K_{\rm od}$=0.192 $\mu$eV/MPa. We can decompose the change of $K_{\rm od}$
 into the contributions from valence bands mixing (red line), and the
VB-CB coupling (blue line). The EPM calculated contribution due
to valence bands mixing is $0.072$ $\mu$eV/MPa, compared with 0.049 $\mu$eV/MPa
from the 6$\times$6 Bir-Pikus model, and the EPM calculated contribution from VB-CB coupling is
0.025 $\mu$eV/MPa compared with 0.036 $\mu$eV/MPa from the 8x8 model Bir-Pikus
model discussed in the Appendix.
 
It is quite surprising that the highly simplified Bir-Pikus model could catch
the change of FSS under external strain rather well, especially, it is known
that the $k$$\cdot$$p$ theory greatly underestimates the FSS in the QDs.
\cite{seguin05} The reason is as follows. The absolute values of FSS are determined by
the combined effects of the strain
distributions, alloy, interfacial effects etc., which can not be captured well
by the continuum theories.  However, these atomistic effects do not change much
if the applied external stress is  not too large. On the other hand, the underlying
crystal structure and electronic structure change coherently under the applied
external stress which break explicitly the $C_{4v}$ symmetry of the system.
Therefore, we expect that the {\it change} of the FSS under stress can be
capture rather well by the Bir-Pikus model, even though the absolute vale of the
FSS could be dramatically underestimated.

\begin{figure}
\includegraphics[width=3.4in]{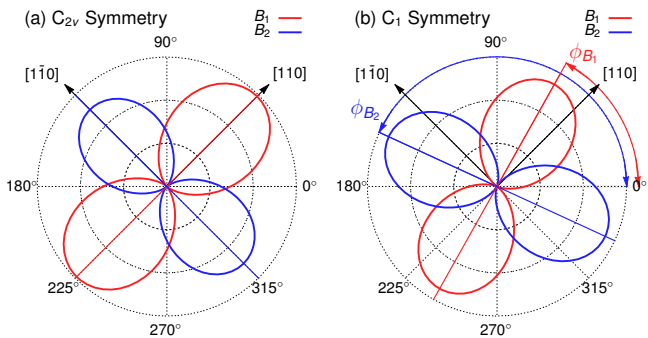}
\caption{\label{fig:dipole} (Color online)
Schematic plot of polar diagrams of (a) dot with $C_{2v}$ symmetry and
(b) dot with $C_{1}$ symmetry.}
\end{figure}

\section{Exciton polarization angle}
\label{sec:polarization}

We now discuss the polarization properties of the two bright exctions using
the above Bir-Pikus model.  The transition dipole matrix elements are given by,
\begin{equation}
\mathcal{M}=\bra{0}\hat{\bf n}\cdot{\bf r} \ket{\Psi_X}\, ,
\end{equation}
where $\hat{\bf n}$ is the polarization vector, and $\Psi_X$ is the exciton
wave function, which is obtained by diagonalize Eq. \ref{eq:exciton}.
The emission intensities of the two bright excitons, passing through a
linear polarizer with an angle $\alpha$ with respect to the [100] axis are
given by,~\cite{Tonin2012}
\begin{eqnarray}
  I_{B_1}(\alpha)
  &=& I_0[\cos(\theta+\alpha)+|\varepsilon_+|\cos(\theta+\phi_\varepsilon-\alpha)]^2
  \, , \nonumber \\
I_{B_2}(\alpha) 
&=&
I_0[\sin(\theta+\alpha)+|\varepsilon_+|\sin(\theta+\phi_\varepsilon-\alpha)]^2\,
,
\label{eq:polarization}
\end{eqnarray}
with $\phi_{\varepsilon}=\mathrm{arg}(\varepsilon_+)$, whereas
2$\theta=-\mathrm{arg}(K_{\rm od})$ is the argument of $K_{\rm od}$, as shown in
Fig.~\ref{fig:dipole}.  Eq.~\ref{eq:polarization} is similar to the
one proposed by Tonin et. al., \cite{Tonin2012} however, the interpretation to the equations is
very different.  In Ref.~\onlinecite{Tonin2012}, $\theta$ is the main elongation
axis orientation with respect to [1$\bar{1}$0] determined by the growing
process, which would not change under the external strain.  In our model, the
polarization angle $\theta$ is determined by the argument of $K_{\rm od}$.  When
stress modifies the exchange integral $K_{\rm od}$ and its argument, the exciton
states rotate in the $x$-$y$ plane accordingly, consistent with the effective
model proposed by the authors in Ref.~\onlinecite{gong11}. 

In the presence of band mixing $\varepsilon_+$, the angle between the two states in the
$x$-$y$ plane will deviate from $\pi/2$ slightly. The polarization angle 
between the two bright exciton states in the $x$-$y$ plane is,
\begin{eqnarray}
\Delta\phi&=&|\phi_{B_2}-\phi_{B_1}|=\frac{\pi}{2}
	  +\arctan\left[-\frac{2|\varepsilon_+|\sin(2\theta+\phi_{\varepsilon})}
	                      {1-|\varepsilon_+|^2} \right] \nonumber      \\ 
&=&\frac{\pi}{2}-2\sin
       [(2\theta+\phi_\varepsilon)] |\varepsilon_+|+O(|\varepsilon_+|^3)\, .
\end{eqnarray}
For dots with $C_{2v}$ symmetry, we have $\phi_{\varepsilon}$=
2$\theta$=$\frac{\pi}{2}$ according to the analysis in Sec. \ref{sec:FSS}, 
and therefore $\Delta\phi={\pi \over 2}$.
The two emission lines are perpendicular to
each other in the $x$-$y$ plane and aligned in the [110] and [1$\bar{1}$0]
directions respectively [See Figure \ref{fig:dipole}~(a)].
This is still true when the dots are under stress along the [110] direction.
For dot with $C_{1}$ symmetry, we have $e_{xx}\neq e_{yy}$ and $e_{zx}\neq
e_{yz}$. Therefore $\phi_{\varepsilon}$ and 2$\theta$  
will deviate from $\frac{\pi}{2}$.
The polarization angles of the two emission lines are,
\begin{equation}
\begin{split}
  \phi_{B_1}&=\arctan\left[-
  \frac{\sin\theta + |\varepsilon_{+}|\sin(\theta+\phi_{\varepsilon})}
       {\cos\theta - |\varepsilon_{+}|\cos(\theta+\phi_{\varepsilon})} \right]
\\ &=-\theta-\sin[(2\theta+\phi_\varepsilon)] |\varepsilon_{+}|
+O(|\varepsilon_+|^2)\, ,
\end{split}
\end{equation}
and
\begin{equation}
\begin{split}
\phi_{B_2}&=\arctan\left[
  \frac{\cos\theta + |\varepsilon_{+}|\sin(\theta+\phi_{\varepsilon})}
       {\sin\theta - |\varepsilon_{+}|\cos(\theta+\phi_{\varepsilon})}\right]
\\ &=\frac{\pi}{2}-\theta+\sin[(2\theta+\phi_\varepsilon)] |\varepsilon_{+}|
+O(|\varepsilon_+|^2)\, ,
\end{split}
\end{equation}
with respect to  [100] direction [See Figure \ref{fig:dipole}~(b)].
In this case, $\Delta\phi \neq$ ${\pi}/{2}$, and the magnitude of the
deviation is proportional to the band mixing parameter $\varepsilon_+$. 

\section{Summary}
\label{sec:summary}

We derive analytically the exciton fine structure splitting 
under the external stress in the self-assembled InAs/GaAs quantum dots 
using the Bir-Pikus model. We find that
the FSS change is mainly due to the strain induced valence bands mixing and 
valence-conduction band coupling. The polarization angle change under strain
is due to the change of the complex phase of the electron-hole off-diagonal exchange integrals. 
The derived theory agrees well with the effective theory and the
empirical pseudopotential calculations, and therefore bridge
the gap between the two theories.

\acknowledgments

LH acknowledges the support from the Chinese National
Fundamental Research Program 2011CB921200, and the
National Natural Science Funds for Distinguished Young Scholars.


\appendix

\section{8$\times$8 Bir-Pikus model}

In the main text of the paper, we neglect the coupling between conduction band (CB) and
valence bands (VB). To include the the VB-CB coupling, we
shall use a $8 \times 8$ model Hamiltonian in the basis set of (e+, e-, HH+,
HH-, LH+, LH-, SO+, SO-),
\begin{widetext}
{ \renewcommand{\arraystretch}{1.5}
\begin{equation}
H=
\left(\begin{array}{cccccccc}
E_g+G & 0 & \sqrt{3}T & 0 & -\sqrt{2}W & T^{\ast} & W & \sqrt{2}T^{\ast} \\
0 & E_g+G & 0 & \sqrt{3} T^{\ast} & T & \sqrt{2}W & \sqrt{2}T & -W \\
\sqrt{3}T^{\ast} & 0 & P+Q+\frac{\Delta}{3}& 0          & -\sqrt{2}S         & R                 & S                & \sqrt{2}R \\
0 & \sqrt{3}T & 0   & p+Q+\frac{\Delta}{3}& R^{\ast}           & \sqrt{2}S^{\ast}  & \sqrt{2}R^{\ast} & -S^{\ast} \\
-\sqrt{2}W & T^{\ast} & -\sqrt{2}S^{\ast} & R          & P-Q+\frac{\Delta}{3}& 0                 & \sqrt{2}Q        & -\sqrt{3}S\\
T & \sqrt{2}W & R^{\ast}          & \sqrt{2}S  & 0                  & P-Q+\frac{\Delta}{3}& \sqrt{3}S^{\ast} & \sqrt{2}Q \\
W & \sqrt{2}T^{\ast} & S^{\ast}          & \sqrt{2}R  & \sqrt{2}Q          & \sqrt{3}S         & P-\frac{2\Delta}{3}& 0         \\
\sqrt{2}T & -W & \sqrt{2}R^{\ast}  & -S         & -\sqrt{3}S^{\ast}  & \sqrt{2}Q         & 0                & P-\frac{2\Delta}{3}
\end{array}\right)\, ,
\end{equation} }
\end{widetext}
where $P$, $Q$, $R$, $S$ are defined in Eq.~\ref{eq:pqrs} in the main text.
The parameters $G$, $W$ and $T$ describe the VB-CB coupling,
\begin{eqnarray}
  G &=& a_{c}(e_{xx}+e_{yy}+e_{zz})\,, \nonumber \\
  W &=& d_{c}e_{xy} \,, \nonumber \\
  T &=& \frac{d_c}{\sqrt{2}}(e_{yz}+ie_{zx}).
\end{eqnarray}
We treat $R$, $S$, $W$ and $T$ as perturbations in the
Hamiltonian and solve the eigenvectors up to the second order of
these parameters. The obtained  hole wave functions are,
\begin{eqnarray}
  \ket{\psi^{v}_{-}}=&\frac{1}{N_{v}}[\ket{\mathrm{hh}+}
+\chi_{\alpha}\ket{\mathrm{lh}+}
+\chi_{\beta}\ket{\mathrm{so}+}
+\varepsilon_{\alpha}\ket{\mathrm{lh}-}
\nonumber\\&+\varepsilon_{\beta}\ket{\mathrm{so}-}
+\eta_{\alpha}\ket{\mathrm{S}+}
+\eta_{\beta}\ket{\mathrm{S}-}
] \, \nonumber \\  
\ket{\psi^{v}_{+}}=&\frac{1}{N_{v}}[\ket{\mathrm{hh}-}
+\varepsilon_{\alpha}^{\ast}\ket{\mathrm{lh}+}
+\varepsilon_{\beta}^{\ast}\ket{\mathrm{so}+}
-\chi_{\alpha}^{\ast}\ket{\mathrm{lh}-}
\nonumber\\&-\chi_{\beta}^{\ast}\ket{\mathrm{so}-}
-\eta_{\beta}^{\ast}\ket{\mathrm{S}+}
+\eta_{\alpha}^{\ast}\ket{\mathrm{S}-}
] \, ,
\end{eqnarray}
with
\begin{equation}
N_{v}=\sqrt{1+|\varepsilon_{\alpha}|^2+|\varepsilon_{\beta}|^2+
|\chi_{\alpha}|^2+|\chi_{\beta}|^2+|\eta_{\alpha}|^2+|\eta_{\beta}|^2}\, .
\end{equation}
The electron wave functions are,
\begin{eqnarray}
  \ket{\psi^{c}_{+}}=&\frac{1}{N_{c}}[
\ket{\mathrm{S}+}
+\tau_{\alpha}^{\ast}\ket{\mathrm{hh}+}
+\tau_{\beta}^{\ast}\ket{\mathrm{hh}-}
+\lambda_{\alpha}^{\ast}\ket{\mathrm{lh}+}
\nonumber\\&+\lambda_{\beta}^{\ast}\ket{\mathrm{so}+}
+l_{\alpha}^{\ast}\ket{\mathrm{lh}-}
+l_{\beta}^{\ast}\ket{\mathrm{so}-}
]\, , \nonumber
\\  \ket{\psi^{c}_{-}}=&\frac{1}{N_{c}}[
\ket{\mathrm{S}-}
-\tau_{\beta}\ket{\mathrm{hh}+}
+\tau_{\alpha}\ket{\mathrm{hh}-}
+l_{\alpha}\ket{\mathrm{lh}+}
\nonumber\\&+l_{\beta}\ket{\mathrm{so}+}
-\lambda_{\alpha}\ket{\mathrm{lh}-}
-\lambda_{\beta}\ket{\mathrm{so}-}
]\, ,
\end{eqnarray}
with
\begin{equation}
N_{c}=\sqrt{1+|\tau_{\alpha}|^2+|\tau_{\beta}|^2+|\lambda_{\alpha}|^2
+|\lambda_{\beta}|^2+|l_{\alpha}|^2+|l_{\beta}|^2}\, .
\end{equation}
Define $a=(X+iY)/\sqrt{2}$ and $b=(X-iY)/\sqrt{2}$, and ignore the
normalization factors, we have,
\begin{eqnarray}
\ket{\psi^{v}_{-}}&=
\ket{(a+\varepsilon_{+}b+\chi_{-}Z+\eta_{\alpha}S)\uparrow
+(\chi_{+}a+\varepsilon_{-}Z+\eta_{\beta}S)\downarrow} \, ,\nonumber \\
\ket{\psi^{c}_{+}}&=
\ket{(S+\tau^{\ast}_{\alpha}a+l^{\ast}_{+}b+\lambda^{\ast}_{-}Z)\uparrow
+(\tau^{\ast}_{\beta}b+\lambda^{\ast}_{+}a+l^{\ast}_{-}Z)\downarrow} \, ,\nonumber\\
\ket{\psi^{v}_{+}}&=
\ket{(b+\varepsilon^{\ast}_{+}a+\chi^{\ast}_{-}Z+\eta_{\alpha}S)\downarrow
-(\chi^{\ast}_{+}b+\varepsilon^{\ast}_{-}Z+\eta_{\beta}S)\uparrow}\, ,\nonumber \\
\ket{\psi^{c}_{-}}&=
\ket{(S+\tau_{\alpha}b+l_{+}a+\lambda_{-}Z)\downarrow
-(\tau_{\beta}a+\lambda_{+}b+l_{-}Z)\uparrow}\, ,
\end{eqnarray}
where
\begin{align}
&\varepsilon_{+}=\frac{\varepsilon_{\alpha}+\sqrt{2}\varepsilon_{\beta}}{\sqrt{3}}\,,\quad
\varepsilon_{-}=\frac{\sqrt{2}\varepsilon_{\alpha}-\varepsilon_{\beta}}{\sqrt{3}}\,,
\nonumber\\ & \chi_{+}=\frac{\chi_{\alpha}+\sqrt{2}\chi_{\beta}}{\sqrt{3}}\,,\quad
\chi_{-}=\frac{-\sqrt{2}\chi_{\alpha}+\chi_{\beta}}{\sqrt{3}}\,,
\nonumber\\ & l_{+}=\frac{l_{\alpha}+\sqrt{2}l_{\beta}}{\sqrt{3}}\,,\quad
l_{-}=\frac{\sqrt{2}l_{\alpha}-l_{\beta}}{\sqrt{3}}\,.
\nonumber\\ & \lambda_{+}=\frac{\lambda_{\alpha}+\sqrt{2}\lambda_{\beta}}{\sqrt{3}}\,,\quad
\lambda_{-}=\frac{-\sqrt{2}\lambda_{\alpha}+\lambda_{\beta}}{\sqrt{3}}\,,
\end{align}
The exchange integral between configurations $\ket{\psi^{v}_{-}\psi^{c}_{+}}$ and $\ket{\psi^{v}_{+}\psi^{c}_{-}}$
\begin{equation}
\begin{split}
K_{\mathrm{od}}
=&\bra{\psi^{v}_{-}\psi^{c}_{+}}\mathcal{K}_{\mathrm{ex}}\ket{\psi^{v}_{+}\psi^{c}_{-}}\\
=&\frac{1}{N^2_{c}N^2_{v}}(\bra{a\uparrow S\uparrow} \mathcal{K}_{\mathrm{ex}} \ket{b\downarrow S\downarrow}
\\&+\varepsilon_{+}
\bra{b\uparrow S\uparrow} \mathcal{K}_{\mathrm{ex}}\ket{b\downarrow S\downarrow}
\\&+\varepsilon_{+}
\bra{a\uparrow S\uparrow} \mathcal{K}_{\mathrm{ex}}\ket{a\downarrow S\downarrow}
\\&+\varepsilon^{2}_{+}
\bra{b\uparrow S\uparrow} \mathcal{K}_{\mathrm{ex}}\ket{a\downarrow S\downarrow}
\\&+\chi_{-}
\bra{Z\uparrow S\uparrow} \mathcal{K}_{\mathrm{ex}}\ket{b\downarrow S\downarrow}
\\&+\chi_{-}
\bra{a\uparrow S\uparrow} \mathcal{K}_{\mathrm{ex}}\ket{Z\downarrow S\downarrow}
\\&+\chi^{2}_{-}
\bra{Z\uparrow S\uparrow} \mathcal{K}_{\mathrm{ex}}\ket{Z\downarrow S\downarrow}
\\&+\varepsilon_{+}\chi_{-}
\bra{b\uparrow S\uparrow} \mathcal{K}_{\mathrm{ex}}\ket{Z\downarrow S\downarrow}
\\&+\varepsilon_{+}\chi_{-}
\bra{Z\uparrow S\uparrow} \mathcal{K}_{\mathrm{ex}}\ket{a\downarrow S\downarrow}
\\&+\tau^2_{\alpha} \bra{a\uparrow a\uparrow} \mathcal{K}_{\mathrm{ex}} \ket{b\downarrow b\downarrow}
\\&+\eta^2_{\alpha} \bra{S\uparrow S\uparrow} \mathcal{K}_{\mathrm{ex}} \ket{S\downarrow S\downarrow}
\\&+\eta_{\alpha} \tau_{\alpha} 
\bra{a\uparrow a\uparrow} \mathcal{K}_{\mathrm{ex}}\ket{S\downarrow S\downarrow}
\\&+\eta_{\alpha} \tau_{\alpha} 
\bra{S\uparrow S\uparrow} \mathcal{K}_{\mathrm{ex}}\ket{b\downarrow b\downarrow}
\\&+\varepsilon_{+}l_{+}\tau_{\alpha}
 \bra{a\uparrow a\uparrow} \mathcal{K}_{\mathrm{ex}}\ket{a\downarrow a\downarrow}
\\&+\varepsilon_{+}l_{+}\tau_{\alpha}
 \bra{b\uparrow b\uparrow} \mathcal{K}_{\mathrm{ex}}\ket{b\downarrow b\downarrow}
\\&+\varepsilon_{+}l_{+}\eta_{\alpha} 
\bra{b\uparrow b\uparrow} \mathcal{K}_{\mathrm{ex}}\ket{S\downarrow S\downarrow}
\\&+\varepsilon_{+}l_{+}\eta_{\alpha} 
 \bra{S\uparrow S\uparrow} \mathcal{K}_{\mathrm{ex}}\ket{a\downarrow a\downarrow}
 \\&+\chi_{-}\lambda_{-}\tau_{\alpha}
 \bra{Z\uparrow Z\uparrow} \mathcal{K}_{\mathrm{ex}}\ket{a\downarrow a\downarrow}
\\& +\chi_{-}\lambda_{-}\tau_{\alpha}
 \bra{b\uparrow b\uparrow} \mathcal{K}_{\mathrm{ex}}\ket{Z\downarrow Z\downarrow}
\\&+\chi_{-}\lambda_{-}\eta_{\alpha} 
\bra{Z\uparrow Z\uparrow} \mathcal{K}_{\mathrm{ex}}\ket{S\downarrow S\downarrow}
\\&+\chi_{-}\lambda_{-}\eta_{\alpha} 
 \bra{S\uparrow S\uparrow} \mathcal{K}_{\mathrm{ex}}\ket{Z\downarrow Z\downarrow}
 \\&+(\varepsilon_{+}l_{+})^2
\bra{b\uparrow b\uparrow} \mathcal{K}_{\mathrm{ex}}\ket{a\downarrow a\downarrow}
\\& +(\chi_{-}\lambda_{-})^2
 \bra{Z\uparrow Z\uparrow} \mathcal{K}_{\mathrm{ex}}\ket{Z\downarrow Z\downarrow}
\\&+\ldots)
\end{split}
\label{eq:Kex}
\end{equation}
There are 1024 terms in total in Eq.~\ref{eq:Kex}, and we list only important
terms. If there is no band mixing, e.g., in the
dots with $C_{4v}$ and above symmetry, only the first term in
Eq.~(\ref{eq:Kex}) exists, which is exact zero for the high symmetry dots
as discussed in the main text.  
If we consider only the valence bands mixing (the 6x6 model in the main text
of the paper), only the first 9 terms in  Eq.~(\ref{eq:Kex}) exist. Other
terms are due to VB-CB coupling.
Because in the InAs/GaAs QDs, the hole-hole Coulomb interactions 
$J_{\mathrm{hh}}=\bra{h\uparrow h\uparrow} \mathcal{K}_{\mathrm{ex}}
\ket{h\downarrow h\downarrow}$ are very
close to the electron-electron Coulomb interactions  
$J_{\mathrm{ee}}=\bra{e\uparrow e\uparrow} \mathcal{K}_{\mathrm{ex}} \ket{e\downarrow e\downarrow}$
and  electron-hole Coulomb interactions  
$J_{\mathrm{eh}}=\bra{e\uparrow e\uparrow} \mathcal{K}_{\mathrm{ex}}
\ket{h\downarrow  h\downarrow}$,
where $h$=$a$ or $b$ or $Z$, are the hole orbitals and $e=S$ is the electron orbital,
we can simplify Eq.~(\ref{eq:Kex}) as
\begin{equation}
\begin{split}
K_{\mathrm{od}}=&\frac{1}{N^2_{c}N^2_{v}} [ (\kappa+i\delta) +2\varepsilon_{+} K 
+2\chi_{-} (\mu+i\nu)
\\&+2\varepsilon_{+}\chi_{-} (\mu-i\nu)
+\varepsilon^2_{+} (\kappa-i\delta) +\chi^2_{-} K_{z}
\\&+(\varepsilon_{+}l_{+}+\chi_{-}\lambda_{-}+\tau_{\alpha})^2 J_ {\mathrm{hh}}
\\&+2(\varepsilon_{+}l_{+}+\chi_{-}\lambda_{-}+\tau_{\alpha})\eta_{\alpha}J_ {\mathrm{eh}}
\\&+\eta^2_{\alpha}J_{\mathrm{ee}}
+\ldots]
\end{split}
\end{equation}
In typical InAs/GaAs QD, $J_{\mathrm{ee}}\approx J_{\mathrm{eh}}\approx J_{\mathrm{hh}}\stackrel{\mathrm{def}}{=} J$.
Additionally  $|\varepsilon_{+}|,|\chi_{-}|,l_{+},\lambda\ll 1$, and $\kappa,\delta,\mu,\nu \ll K, K_{z}\ll J$,
therefore $K_{\mathrm{od}}$ can be further simplified as,
\begin{equation}
K_{\mathrm{od}}\approx (\kappa+i\delta) + 2\varepsilon_{+} K+(\varepsilon_{+}l_{+}+\chi_{-}\lambda_{-}+\tau_{\alpha}+\eta_{\alpha})^2 J
\label{eq:Kex8}
\end{equation}
%
The first two terms in the above equation has been obtained from the
6$\times$6 model, whereas the last term is due to VB-CB coupling. 
As we shall see later, though the VB-CB coupling
is much weaker than the valence bands mixing, the contribution to the FSS
change is still significant, 
because $J \sim$ 20 meV is much larger than $K \sim$ 100 $\mu$eV. Therefore, 
we keep higher order terms of VB-CB band mixing.

Let us denote $E'_g=E_g+G-P-Q-\Delta/3$ and $E''_g=E_g+G-P+2Q+2\Delta/3$, where
$E_g \approx$ 1 eV is the single-particle band gap of the strained InAs. After
tedious but straightforward derivation, we have,
\begin{align}
\tau_{\alpha}+\eta_{\alpha} &\approx 0 \, ,\\
\varepsilon_{+} &\approx \frac{(9Q+\Delta)R^{\ast}}{2\sqrt{3}Q\Delta}\, , \\
l_{+} &\approx \frac{\sqrt{3}T^{\ast}}{E'_g}\, ,\\
\chi_{-} &\approx  \frac{S^{\ast}}{\sqrt{3}Q}\, ,\\
\lambda_{-} &\approx \frac{\sqrt{3}W}{E''_g}\, .
\end{align}
Here we have used the fact that $E'_g \sim$ $E''_g \gg$ $Q$, $\Delta$.
By using the relationship $W=i(R-R^{\ast})d_{c}/(2d_{v})$ and
$T=iSd_{c}/d_{v}$, we obtain,
\begin{equation}
\begin{split}
&\varepsilon_{+}l_{+}+\chi_{-}\lambda_{-}+\tau_{\alpha}+\eta_{\alpha}
\\ \approx &
i\frac{d_c}{d_v}\frac{S^{\ast}}{2Q}
\left[\frac{R-R^{\ast}}{E''_g}-\frac{R^{\ast}}{E'}\left(\frac{9Q}{\Delta}+1\right)
\right]
\end{split}
\end{equation}
It is easy to show that  
${d (\tau_{\alpha} + \eta_{\alpha})}/{dp}$, ${d
    l_{+}}/{dp}$, ${d\chi_{-}}/{dp}$, ${d\lambda_{-}}/{dp} \ll$
  ${d\varepsilon_{+}}/{dp}$, which has been obtained in Eq.~\ref{eq:dvarepsilon}. 
For the stress along the [110] direction, we have
\begin{equation}
\begin{split}
\frac{dK_{\mathrm{od}}}{dp}&\approx2\frac{d\varepsilon_{+}}{dp}K+
2(\varepsilon_{+}l_{+}+\chi_{-}\lambda_{-}+\tau_{\alpha}+\eta_{\alpha})l_{+}\frac{d\varepsilon_{+}}{dp}J
\\&\approx -i\frac{9Q+\Delta}{4\sqrt{3}Q\Delta}d_v S_{44}K \\
& + i\frac{1}{8}\left(\frac{d_c}{d_v}\frac{S^{\ast}}{E'_g}\frac{9Q+\Delta}{Q\Delta}\right)^2
R^{\ast}d_v S_{44} J \, ,
\end{split}
\label{eq:slope8}
\end{equation}
where the first term has been obtained from the 6$\times$6 model in the main text,
and the second term is due to the VB-CB coupling.
In typical InAs/GaAs QDs, $E'_g\sim$ 1 eV and $E''_g\sim$ 2 eV,
$Q\sim$ 0.23 eV, $\Delta\sim$ 0.34 eV and $|S|\sim$ 0.065 eV.  
The electron-hole Coulomb interactions $J\sim$
20 meV, and dark-bright splitting energy 2$K\sim$200 $\mu$eV.
Using the parameters given in Table I, we have 
${d\varepsilon_{+}}/{dp} \approx i 2.4 \times 10^{-4}$ MPa$^{-1}$.
Finally, we obtain
${dK_{\mathrm{od}}}/{dp} \approx i(0.049+0.036)$=$i$0.084 $\mu$eV/MPa, which
is in a reasonable good agreement with the EPM calculations and the experimental values. 



\end{document}